\documentclass[prl,twocolumn,10pt,tightenlines,preprint]{revtex4}
\usepackage{amsfonts}
\usepackage{graphicx}

\begin{document}

\title{Comment on \textit{Deterministic Single-Photon Source for Distributed
Quantum Networking} by Kuhn, Hennrich, and Rempe}
\author{H. J. Kimble}
\affiliation{Norman Bridge Laboratory of Physics 12-33\\
California Institute of Technology\\
Pasadena, CA 91125 USA}
\maketitle

A recent article \cite{kuhn02} describes an experiment to generate
single photons within the setting of cavity QED. The authors claim
that \textquotedblleft a sequence of single photons is emitted on
demand\textquotedblright\ and that their results represent
\textquotedblleft the realization of an intrinsically reversible
single-photon source.\textquotedblright\ Although their work is
certainly an advance towards these goals, unfortunately the
observational evidence reported in Ref. \cite{kuhn02} does not
support the principal claims of the demonstration of a
deterministic source for single photons, nor of emission that is
intrinsically reversible as is required for the transfer of
quantum states over a network. The underlying difficulties are (1)
the random arrival of atoms into the interaction region means that
photons are emitted at random and not ``on demand,'' (2) the
corresponding fluctuations in atom number lead to a photon stream
that is super-Poissonian, and (3) the stochastic character of
atomic trajectories through the cavity mode produces unknown
variations in the amplitude and phase of the emitted field, so
that the source is not ``intrinsically reversible.''

I begin by examining the data presented in Figure 4 of Ref.
\cite{kuhn02}, which displays the second-order intensity
correlation function $g_{D_{1},D_{2}}^{(2)}(\tau )$ for the
cross-correlation of photoelectric counting events from two
detectors $(D_{1},D_{2})$ as a function of time separation $\tau
$. Somewhat surprisingly, $g_{D_{1},D_{2}}^{(2)}(\tau )\geq 1$,
and in particular, $g_{D_{1},D_{2}}^{(2)}(0)\simeq 1$, so that the
inferred photon statistics are \textit{super-Poissonian}
$\left\langle \Delta n^{2}\right\rangle >\left\langle
n\right\rangle $ \cite{m-w-95}. This is in marked contrast to the
behavior required for an \textquotedblleft
on-demand\textquotedblright\ single-photon source, for which $%
g^{(2)}(0)\simeq 0$ with \textit{sub-Poissonian} photon statistics $%
\left\langle \Delta n^{2}\right\rangle <\left\langle n\right\rangle $ \cite%
{m-w-95}. The authors attribute this disparity to detection events other
than those arising from photons emitted from the cavity. \textit{However, I
emphasize that }$g_{D_{1},D_{2}}^{(2)}(\tau )$\textit{\ would maintain the
same form as in Fig. 4 of Ref. \cite{kuhn02} even if the \textquotedblleft
relatively large noise contribution\textquotedblright\ from background light
were eliminated altogether.}

\begin{figure}[b]
\includegraphics[width=8.5cm]{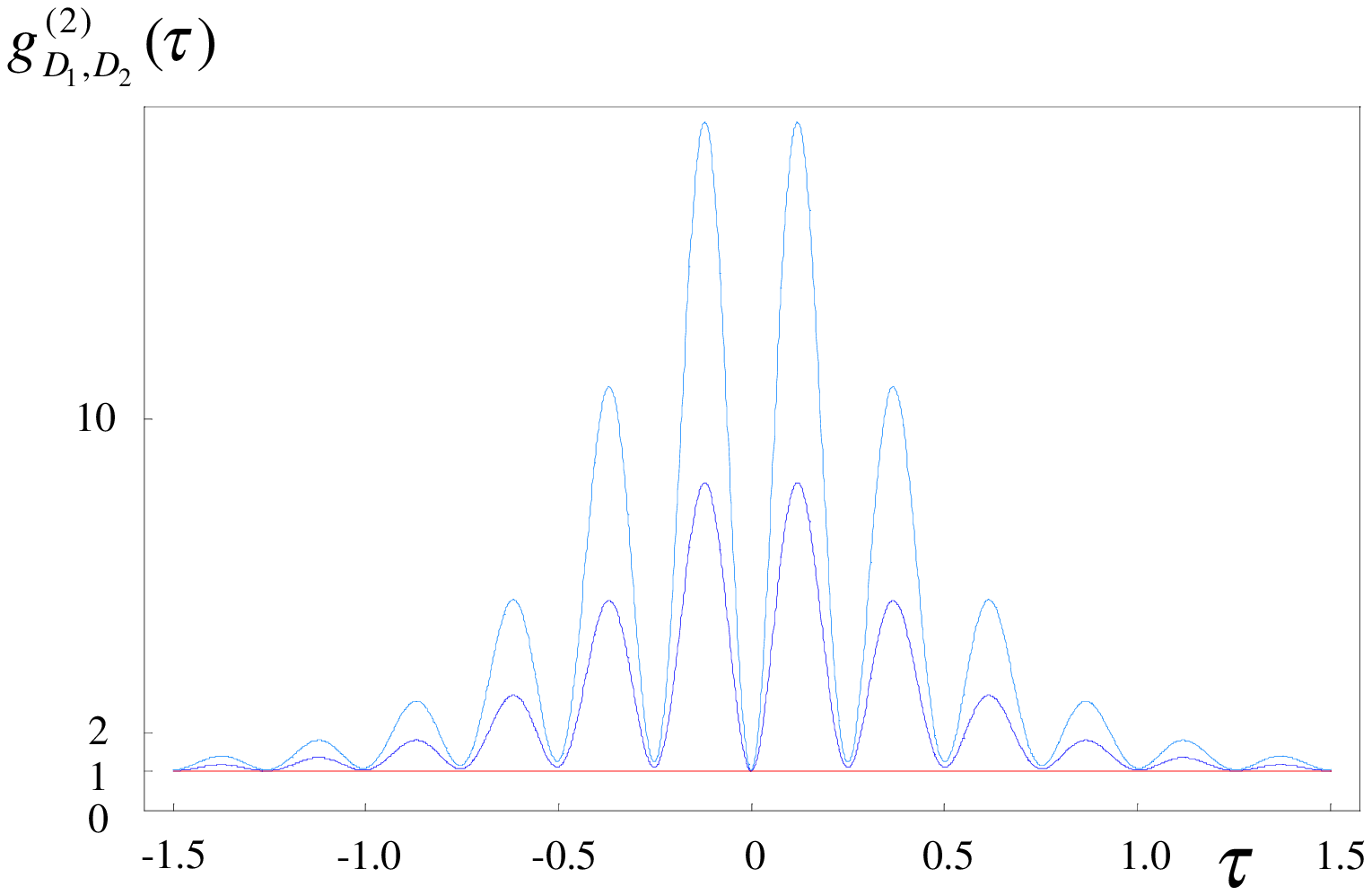}
\caption{\label{g2tau}Intensity correlation function
$g_{D_{1},D_{2}}^{(2)}(\tau)$ versus time delay $\tau$ for the
fluorescent light from a beam of atoms with average number
$\bar{N}=0.1$ atoms. The line at $g_{D_{1},D_{2}}^{(2)}(\tau)=1$
represents the Poisson limit for coherent light of the same mean
counting rates at $(D_{1},D_{2})$. Time delay $\tau$ is measured
in units of the transit time $t_{0}$. The generalized Rabi
frequency $\Omega^{\prime}t_{0}=25$ and transverse decay rate
$\beta t_{0}=0.1$. The lower trace is for background to signal
ratio $=0.5$, while the upper trace has no background
\cite{kimble78}.}
\end{figure}

To illustrate this point, consider the well-studied system of
resonance fluorescence from a single two-state atom
for which the photon statistics are described by $g_{A}^{(2)}(\tau )$, with $%
g_{A}^{(2)}(0)=0$ \cite{m-w-95}. If observations are made not for
a single atom but rather for an interaction volume with a
stochastic variation in atom number $N$, the resulting intensity
correlation function $g_{D_{1},D_{2}}^{(2)}(\tau )$ is of a
markedly different form, as illustrated in Figure \ref{g2tau}. \textit{%
Significantly, Figure \ref{g2tau} reproduces the essential
characteristics of Figure 4 in Ref. }\cite{kuhn02}\textit{,
including that the light is super-Poissonian} \cite{small}. The
commonality of these two figures arises because of fluctuations in
the number of ``source'' atoms about which there is no independent
knowledge. In this setting, the observation of sub-Poissonian
\textit{photon statistics} requires sub-Poissonian \textit{atom
statistics}, with $g_{D_{1},D_{2}}^{(2)}(0)<1$ in direct
correspondence to the reduction $Q_{A}\equiv
\frac{\overline{(\Delta N)^{2}}-\overline{N}}{\overline{N}}<0$
\cite{kimble78}. Strategies to achieve $Q_{A}<0$ include
conditional detection as employed in the original experiment of
Short and Mandel \cite{m-w-95,numbers} and atom trapping within
the cavity.

In addition to fluctuations in arrival time and atom number, the experiment of Ref. \cite%
{kuhn02} suffers from a lack of atomic localization with respect to the
spatially varying coupling coefficient $g(\vec{r})$ due to unknown atomic
motion through the cavity mode. As as result, the output pulse shapes and
phases for photon emissions vary in a random fashion and are not
controllable \cite{duan02}. For the parameters of Ref. \cite{kuhn02}, the
transverse velocity of a typical atom carries it across distances $\pm \frac{%
\lambda }{4}$ along the cavity axis, leading to (random)
variations $\sim \pm \frac{\pi }{2}$ in the phase of the emitted
field. \textit{Hence, photon emissions are reversible only in the
sense that the emitted field is returned to the very same atom
that gave rise to the
emission within a time short compared to the atomic transit time }$%
t_{0}\equiv \frac{2w_{0}}{v_{z}}\simeq 35\mu $\textit{s.}
Moreover, reversible transmission to a second atom-cavity system
requires knowledge of the actual time of the initial emission, as
well as an ``event'' ready atom at the remote location. Neither of
these capabilities follows from the experiment reported in Ref.
\cite{kuhn02}.


\begin{thebibliography}{9}
\bibitem{kuhn02} A. Kuhn, M. Hennrich, and G. Rempe, Phys. Rev. Lett.
\textbf{89}, 067901 (2002).

\bibitem{m-w-95} \textit{Optical Coherence and Quantum Optics}, L. Mandel
and E. Wolf (Cambridge University Press, 1995), Section 15.6.

\bibitem{kimble78} H. J. Kimble, M. Dagenais, and L. Mandel, Phys. Rev.
\textbf{A18}, 201 (1978), Eq. (31).

\bibitem{small} Remaining quantitative differences (e.g., around $\tau =0$)
could be resolved by a more detailed model for the internal atomic states
and the pumping and recycling mechanisms.

\bibitem{numbers} Note that the flux of atoms through the cavity in Ref.
\cite{kuhn02} is completely asynchronous with respect to the
driving lasers. Furthermore, the mean number of photoelectric
events per atom per pumping cycle is $\sim 0.04\ll 1$. This low
efficiency and the high background rate severely limit any
sub-Poissonian effect that might be observed via conditional
atomic detection \cite{m-w-95}.

\bibitem{duan02} L.-M. Duan, A. Kuzmich, and H. J. Kimble, quant-ph/0208051.

\end{thebibliography}
\end{document}